%% file: main.tex
\documentclass[conference]{IEEEtran}
\IEEEoverridecommandlockouts

\usepackage{cite}
\usepackage{amsmath,amssymb,amsfonts}
\usepackage{algorithmic}
\usepackage{graphicx}
\usepackage{textcomp}
\usepackage{xcolor}
\usepackage{orcidlink}
\usepackage{listings}
\usepackage{tcolorbox}
\usepackage{dirtytalk}
\usepackage{multirow}

\input{macros.tex}

\begin{document}
\title{\chatgpt{} and Software Testing Education: \\Promises \& Perils
}

\author{
\IEEEauthorblockN{ Sajed Jalil \orcidlink{0000-0003-1249-2113}\\
sjalil@gmu.edu
}
\and
\IEEEauthorblockN{ Suzzana Rafi \orcidlink{0000-0002-3244-5800}\\
srafi@gmu.edu}
\and
\IEEEauthorblockN{ Thomas D. LaToza \orcidlink{0000-0002-9564-3337}\\
tlatoza@gmu.edu}
\and
\IEEEauthorblockN{ Kevin Moran \orcidlink{0000-0001-9683-5616}\\
kpmoran@gmu.edu}
\and
\IEEEauthorblockN{ Wing Lam \orcidlink{0000-0003-2243-1218}\\
winglam@gmu.edu}
\linebreakand 
\IEEEauthorblockA{\textit{Department of Computer Science} \\
\textit{George Mason University}\\
Fairfax, USA \\
}
} 

\maketitle
\thispagestyle{plain}
\pagestyle{plain}

\begin{abstract}
Over the past decade, predictive language modeling for code has proven to be a valuable tool for enabling new forms of automation for developers. 
More recently, we have seen the advent of general purpose ``large language models", based on neural transformer architectures, that have been trained on massive datasets of human written text, which includes code and natural language. 
However, despite the demonstrated representational power of such models, interacting with them has historically been constrained to specific task settings, limiting their general applicability. 
Many of these limitations were recently overcome with the introduction of \chatgpt{}, a language model created by OpenAI and trained to operate as a conversational agent, enabling it to answer questions and respond to a wide variety of commands from end users.

The introduction of models, such as \chatgpt{}, has already spurred fervent discussion from educators, ranging from fear that students could use these AI tools to circumvent learning, to excitement about the new types of learning opportunities that they might unlock. 
However, given the nascent nature of these tools, we currently lack fundamental knowledge related to how well they perform in different educational settings, and the potential promise (or danger) that they might pose to traditional forms of instruction. 
As such, in this paper, we examine how well \chatgpt{} performs when tasked with answering practice questions in a popular software testing curriculum. 
We found that given its current capabilities, \chatgpt{} is able to respond to 77.5\% of the questions we examined and that, of these questions, it is able to provide correct or partially correct \textit{answers} in \answerAccuracyWithPartial{}\% of cases, provide correct or partially correct \textit{explanations} of answers in \explanationAccuracyWithPartial{}\% of cases, and that prompting the tool in a shared question context leads to a marginally higher rate of correct answers and explanations.
Based on these findings, we discuss the potential promises and perils related to the use of \chatgpt{} by students and instructors.
\end{abstract}

\begin{IEEEkeywords}
\chatgpt{}, testing, education, case study
\end{IEEEkeywords}

\section{Introduction}

Language modeling of code has been an important topic in software engineering research since the promise of modeling code was first illustrated by Hindle et al.~\cite{Hindle:ICSE'12}.
As techniques for language modeling improved, researchers began to utilize Deep Learning (DL) architectures to learn rich, hierarchical representations of code that could then be used for various downstream tasks~\cite{White:MSR'15}. 
In parallel, the machine learning and natural language processing communities began building large-scale models centered around a specific type of neural architecture, the transformer~\cite{Vaswani:NeurIPS'17,Brown:NeurIPS'20,devlin-etal-2019-bert}, trained on massive datasets of text. Experiments illustrated the representational power of both these large language models (LLMs), and language models tailored specifically for code~\cite{AlphaCodeEval2022,feng-emnlp'20,allamanis2018learning,Chen:arxiv'21}. However, such models were largely constrained to\Space{ highly} specific task settings and did not provide for \textit{natural} forms of interaction with end users -- until recently.

In late 2022, OpenAI introduced \chatgpt{}~\cite{chatgptwebsite}, an AI tool built on top of existing LLMs, that enabled interaction through a conversational interface. 
To enable this type of interaction\Space{ with the model}, OpenAI made use of reinforcement learning from human feedback, refining methods from past work on InstructGPT~\cite{ouyang2022training}, which trained LLMs with both unsupervised data and with supervision in the form of task instruction. 
In essence, the model was initially trained on real, human text-based conversations, then learned to refine its responses based on feedback from human evaluators that rated the quality of answers in a reinforcement learning setting. 
This process proved very successful in creating an interface where users could easily access the latent ``knowledge'' of LLMs.

Given the ease of interaction and the seemingly vast amount of knowledge contained within the model, vigorous discussion arose in academic communities about the potential dangers and opportunities of such a tool for both students and instructors. 
Sentiments ranged from fear that students would use these models to circumvent learning material, to excitement about the new types of learning, assignments, and automation a tool, such as \chatgpt{}, could introduce into various levels of academia. 
However, to make informed decisions about how to use (or not use) \chatgpt{} in various educational settings, we must first have a thorough understanding of its capabilities, strengths, and weaknesses. 

In this paper, we aim to study and understand the capabilities of \chatgpt{} in the context of a traditional software testing course. 
As such, we conduct a comprehensive empirical study, tasking \chatgpt{} with answering questions from five chapters of a popular software testing textbook~\cite{ammann2016introduction}, and thoroughly vetting the results across multiple dimensions. 
We aim to learn (i) how often \chatgpt{} is correct in answering questions, (ii) how often it can fully and accurately explain its answers, (iii) how different ways of asking questions to \chatgpt{} can affect its ability to provide correct responses\OurComment{how the non-determinism in \chatgpt{}'s response can affect its ability to provide correct responses}, and (iv) whether \chatgpt{}'s expressed confidence provides bearing on the correctness of its answers. 
We find the current capabilities of \chatgpt{} allow it to properly respond to 77.5\% of the testing questions we examined. For the questions to which \chatgpt{} was able to respond, it provides correct or partially correct \textit{answers} in \answerAccuracyWithPartial{}\% of cases, and provides correct or partially correct \textit{explanations} in \explanationAccuracyWithPartial{}\% of cases We also found that prompting the model in a shared context, where similar questions are asked together, lead to marginally better answers, and the tool's claimed confidence level seems to have little bearing on the correctness of the answers. 
Based on these findings, we discuss the potential promises and perils related to the use of \chatgpt{} in software testing courses.

In summary, the contributions of this paper are as follows:

\begin{itemize}
    \item{A manually-vetted dataset of \chatgpt{}'s answers to \numAllQuestions{} questions from five chapters of a popular software testing textbook. Our dataset includes three responses from \chatgpt{} for each question.}
    \item{A thorough analysis of these answers that examines how often \chatgpt{} is correct and when it is able to properly explain a given answer.}
    \item{An investigation into two prompting strategies, and their effect on answer and explanation correctness, as well as an analysis of whether \chatgpt{}'s proclaimed confidence level impacts answer/explanation correctness.}
    \item{An online appendix~\cite{studygithub,zenodo}, that includes our data, analysis code, and experimental infrastructure to facilitate replicability and future work on applications of \chatgpt{} to various topics in computer science education.}
\end{itemize}

\Space{
\KEVIN{This should go in the study methodology section, not the intro. This could probably be dropped entirely.}\Fix{Text that can be used in intro to explain different contexts of using \chatgpt{}:}
This example is a multi-part question -- it has two parts that are used for this study\Space{ as we only dealt with questions that are in the student solution}. 
Both of the \subquestion{}s are given to \chatgpt{} at once along with the common context. 
In this study, the context of the prompt given to \chatgpt{} was divided into two groups - shared and separate. 
\begin{itemize}
    \item Shared Context is when all the \subquestion{}s of a multi-part question are asked in a single context one after another.
    \item Separate Context is when all the \subquestion{}s are asked in separate contexts. In each context, the common part of the question is given along with each \subquestion{}.
\end{itemize}
If a question does not have multiple parts, then shared and separate contexts are the same. 
We evaluate how \chatgpt{} performs under these two different types of contexts in RQ1 (Section~\ref{sec:eval:rq1}).
}

\section{Background}
\label{sec:background}

\chatgpt{}~\cite{chatgptwebsite} offers a machine learning model designed to engage in conversations with the user. 
It provides responses to questions asked in a prompt and is able to respond to follow-up questions and correct itself.

To investigate the applicability of ChatGPT to answer questions commonly found in a software testing curriculum, we pose questions from the first five chapters of the textbook by \AO~\cite{ammann2016introduction}.
This book has been used in undergraduate and graduate software testing classes at George Mason University and is used by hundreds of organizations as a source of software testing knowledge. Our study uses:

\begin{itemize}
    \item Chapters 1 and 2, which discuss software faults, errors, and failures
    \item Chapter 3, which discusses the Reachability, Infection, Propagation, and Revealability (RIPR) model
    \item Chapter 4, which discusses Test Driven Development (TDD) and continuous integration 
    \item Chapter 5, which discusses coverage criterion and subsumption
\end{itemize} 
The remainder of this section provides a brief description of the course content found in each chapter.

\subsection{Chapters 1\& 2 - Fault, Error and Failure} 

A \emph{fault} is a static defect in the software. An \emph{error} is an incorrect internal state, which is composed of a program counter and the live variables at that program counter location, and a \emph{failure} is an external, incorrect behavior with respect to the requirements or the description of the expected behavior~\cite{ammann2016introduction}.

\subsection{Chapter 3 - RIPR Model}

According to \AO~\cite{ammann2016introduction}, there are four conditions that are needed for a failure to be observed. These conditions together are called the RIPR model. First, a test needs to reach the location of the defective line of code – that is \emph{Reachability}. After the fault is executed, it leads to an incorrect program state, which is called \emph{Infection}. The infection must spread to an incorrect final state – that is \emph{Propagation} and finally, the incorrect portion of the final state must be observable by the tester – that is \emph{Revealability}. 

\subsection{Chapter 4 - CI \& TDD}

\AO~\cite{ammann2016introduction} describe Continuous Integration (CI) as the process that begins with a developer using a fresh development environment, obtaining the code under test and test code\Space{ from a project repository}, building the code, and running the tests. 
After finalizing changes to code, the changes start the CI process, where a fresh environment rebuilds the code and reruns the tests. 
This process helps developers quickly identify any failures their changes may have introduced\OurComment{ and also keeps the whole team informed of any differing design choices}.

Test Driven Development (TDD) is a methodology for creating software in which tests are written before the code under test. 
The methodology is based on repeating a short development cycle, including creating a test, running it to confirm that it fails, writing code under test to make the test pass, creating more tests, and improving the code under test to make more tests pass. 
TDD aims to produce clean and failure-free code by writing only code under test to make tests pass.

\subsection{Chapter 5 - Coverage Criterion and Subsumption}
\label{sec:background:example}

\Space{The topic of coverage criterion and subsumption is covered in Chapter 5~\cite{ammann2016introduction}.}
A coverage criterion is a rule or collection of rules that impose test requirements on a test set. 
A coverage criterion C1 subsumes C2 if and only if every test set that satisfies criterion C1 also satisfies C2. 

\section{Study Setup}

To investigate the promises and perils of using ChatGPT for software testing education, we study the following\Space{ four} research questions (RQs):

\begin{itemize}
    \item \RQONE
    \item \RQTWO
    \item \RQTHREE
    \item \RQFOUR
\end{itemize}

\begin{table}
\vspace{-2em}
\caption{\label{tab:questions} Overview of the questions in our study.}
\postcaptionsCodespace
\begin{tabular} {  c  c  c  c  c  c }
\textbf{Chapter} & \textbf{Question} & \textbf{Sub-Question{}} & \textbf{Code} & \textbf{Concept} & \textbf{Both}\\
\hline
1 & 5.2.a &	\checkmark & & & \checkmark\\
1 & 5.2.b &	\checkmark & & & \checkmark\\
1 & 5.2.c &	\checkmark & & & \checkmark\\
1 & 5.2.d &	\checkmark & & & \checkmark\\
1 & 5.2.e &	\checkmark &  & \checkmark & \\
1 & 5.4.a &	\checkmark & & & \checkmark\\
1 & 5.4.b &	\checkmark & & & \checkmark\\
1 & 5.4.c &	\checkmark & & & \checkmark\\
1 & 5.4.d &	\checkmark & & & \checkmark\\
1 & 5.4.e &	\checkmark & & \checkmark & \\
1 & 7.2.a &	\checkmark & & & \checkmark\\
1 & 7.2.b &	\checkmark & & & \checkmark\\
1 & 7.2.c &	\checkmark & & & \checkmark\\
1 & 7.2.d &	\checkmark & & & \checkmark\\
1 & 7.2.e &	\checkmark & & \checkmark & \\
1 & 7.3.a &	\checkmark & & & \checkmark\\
1 & 7.3.b &	\checkmark & & & \checkmark\\
1 & 7.3.c &	\checkmark & & & \checkmark\\
1 & 7.3.d &	\checkmark & & & \checkmark\\
1 & 7.3.e &	\checkmark & & \checkmark & \\
2 & 1 &	 & & \checkmark & \\
3 & 4 &	 & \checkmark & \\
3 & 5 &	 & & \checkmark & \\
3 & 9.a &	\checkmark & \checkmark & \\
3 & 9.b &	\checkmark & & \checkmark & \\
3 & 9.c &	\checkmark & \checkmark & \\
3 & 9.d &	\checkmark & \checkmark & \\
3 & 9.e &	\checkmark & \checkmark & \\
4 & 1 &	 & \checkmark &  & \\
5 & 1.a & \checkmark & & \checkmark & \\
5 & 1.b & \checkmark & & \checkmark & \\
\hline
\textbf{Count} & \textbf{\numAllQuestions{}} & \textbf{\numSubQuestions{}} & \textbf{\codeQuestion{}} & \textbf{\conceptualQuestion{}} & \textbf{\combinedQuestion{}}\\
\end{tabular}
\vspace{-2em}
\end{table}

\subsection{Dataset}
\label{sec:study:dataset}

Our dataset contains questions from the widely used software testing book by \AO~\cite{ammann2016introduction}.
In the context of this study, we use all the textbook questions in Chapters 1 to 5 that have solutions available on the book's official website~\cite{booksolution}.
These solutions are made publicly available to help students learn.
We omitted questions that do not have student solutions, as publishing our results might expose answers that the authors of the book do not intend to make public.
Our study is limited to the first five chapters of the book, which emphasize topics taught in most introductory software testing courses.\Space{These topics are described in Section~\ref{sec:background}.}
Often, our selected questions have \emph{\subquestion{}s}, which break down a more comprehensive question into smaller parts. 
For example, for a given code snippet, there might be multiple related \subquestion{}s that each ask about properties of the snippet. In our study, for simplicity, we refer to all questions and \subquestion{}s as \textit{questions} - given that we treat them all as having equal importance.

We collected a total of 40 exercise questions that meet our requirements.
After manual inspection, we identified nine questions that ask for material that is impossible for \chatgpt{} to generate, as it is capable of generating only text-based responses. 
For example, we encountered questions that ask for a screen printout of code execution, a project to be fetched from the internet, or a continuous integration server to be set up. 
Questions with such tasks cannot be fully and correctly answered by \chatgpt{}'s\Space{ current capabilities of providing only} text-based responses.

We removed the nine questions that are outside of \chatgpt{}'s capabilities to correctly answer and report our results on a total of \numAllQuestions{} questions to which \chatgpt{} may give a correct response. 
Of\Space{ the selected} \numAllQuestions{} questions, 
\numQuestionsWSub{} are multi-part questions that collectively contain 27 \subquestion{}s and \numNonSubQuestions{} are independent questions that do not contain any \subquestion{}s.

Table~\ref{tab:questions} lists the characteristics of each question in our dataset. 
We find that six questions ask only for code in their answers, nine ask to explain a concept, and the remaining 16 ask for both code and a concept explanation. 
Our dataset is publicly available on GitHub~\cite{studygithub} and Zenodo~\cite{zenodo}.

\subsection{Data Collection Tool}
During the data collection process for this study, OpenAI~\cite{openai} had not yet made a \chatgpt{} API publicly available, therefore, the number of questions we were able to ask \chatgpt{} through the online interface during a given period of time was rate-limited.
We developed an open-source tool to collect \chatgpt{} responses for the questions in our dataset~\cite{studygithub}.
Our tool automatically queries \chatgpt{}, collects the responses, and waits \tenSeconds{} seconds after receiving an answer before asking the next question.
We determined the length of this delay based on experiments with our automated tool (e.g., trying multiple wait times\Space{ to determine the shortest acceptable by the interface}).
We find that a wait time of \tenSeconds{} seconds provided us the greatest number of query responses.

\subsection{Methodology}
For \textbf{RQ}$_1$, we look to understand how often \chatgpt{} is able to provide correct answers and explanations to our dataset of software testing questions and to determine how \chatgpt{} performs when \subquestion{}s are asked in a single chat context one by one compared to when they are asked in separate contexts. 
We refer to these two ways of asking questions as \emph{shared context} and \emph{separate context}, respectively. 
For \textbf{RQ}$_2$, we study how often \chatgpt{} will give answer-explanation pairs with different degrees of correctness (e.g., correct answer but incorrect explanation)\Space{ using the context that is most likely to give correct answers}.
For \textbf{RQ}$_3$, we aim to analyze how \chatgpt{}'s non-determinism affects its ability to provide correct answers and explanations by posing each question three times and examining any differences in responses.
Lastly, for \textbf{RQ}$_4$, we aim to determine whether \chatgpt{}'s self-reported confidence (which can be collected through a \textit{confidence query}) related to an answer/explanation has a bearing on the correctness of that answer. 
Our findings could be useful for instructors and students to determine the potential utility of a given answer. Below, we define the processes for (1) \textit{shared context queries}, (2) \textit{separate context queries}, (3) \textit{confidence queries}, and (4) \textit{response labeling}.

\subsubsection{\underline{Separate Context Query}}
In \emph{separate} context queries, we treat each of the \numSubQuestions{} \subquestion{}s as an independent question. 
Each \subquestion{} is asked in a separate chat context. 
Combining with the \numNonSubQuestions{} independent questions, a total of \numAllQuestions{} questions are asked for each run. 
To evaluate how separate context compares with shared context, we collect a total of three runs for each question, which results in a total of \numAllQuestionsThreeRuns{} separate context responses.

\subsubsection{\underline{Shared Context Query}}
In this query scenario, sub-questions are all asked in a single \chatgpt{} session during which the context of the conversation is shared (i.e., \chatgpt{} is able to reference an initial prompt or code snippet, and parts of prior sub-questions) as long as the \subquestion{}s refer to the same code or scenario. 
For example, consider the \lastZero{} method in Figure~\ref{fig:lastzero}.
\lastZero{} is supposed to find the last index in an array where a zero occurs. 
One may ask multiple \subquestion{}s based on this code (e.g., give a test that (1) does not execute the fault or (2) executes the fault, but there are no failures). 

\begin{figure}[t]
	\begin{lstlisting}[language=Java,basicstyle=\ttfamily\footnotesize,escapeinside={(*@}{@*)},columns=fixed,xleftmargin=3.5ex]
public static int lastZero (int[] x) {
  for (int i = 0; i < x.length; i++)
    if (x[i] == 0) return i;
  return -1;
}
    \end{lstlisting}
    \precaptionspace
	\caption{Code snippet for \lastZero{}.}
	\label{fig:lastzero}
	\postcaptionsCodespace
\end{figure}

In the shared context query, we provide the implementation of \lastZero{} and ask the first \subquestion{} in one chat context. 
After we obtain the response, we continue to ask the next \subquestion{} in the same, \emph{shared} context. 
This process is repeated until all \subquestion{}s are asked before the context is destroyed. 
For the next question with multiple \subquestion{}s, we then open another chat context.

Overall, we obtain 81 responses from running each of the \numSubQuestions{} \subquestion{}s three times and 12 responses from running each of the \numNonSubQuestions{} independent questions three times\Space{ that do not have \subquestion{}s}. 
For questions with no \subquestion{}s, shared context is the same as separate context, as such, the responses to the four independent questions are identical across the datasets for both the shared and separate contexts, with only the 81 responses to the 27 sub-questions differing. 
In \textbf{RQ}$_1$, we compare responses for which both shared context and separate context exists, i.e., 81 responses.
We find that shared context is more likely to give correct responses than separate context.
In \textbf{RQ}$_2$-\textbf{RQ}$_4$ we use the responses from shared context -- the 81 shared context responses and the 12 shared/separate context responses.

\subsubsection{\underline{Confidence Query}}
To determine how confident \chatgpt{} is in its answer for RQ4, we ask the following question to \chatgpt{} after each of its responses -- \emph{``\confidenceQuestion{}''}. 
To answer \textbf{RQ}$_1$-\textbf{RQ}$_3$, we asked each question 3 times. However, to simplify our data collection process, we only collected responses to confidence queries the first time that a question was posed to \chatgpt{} in the shared context setting.
We recorded the following replies from \chatgpt{}: Highly confident, Very Confident, Confident, and Reliable.
Note that we do not attempt to rank \chatgpt{}'s confidence replies, as their relative ranking is unclear (e.g., whether ``Highly confident'' is more or less confident than ``Very confident'').
We attempted to resolve the ranking of these confidence replies by asking \chatgpt{} to provide a relative ranking, but \chatgpt{} was not able to provide a conclusive response. 
Therefore, for \textbf{RQ}$_4$, we simply study the likelihood for \chatgpt{} to give a correct explanation or answer given a specific category of confidence reply.

\subsubsection{\underline{Response Labeling}}
\label{sec:lackinfo:example}
To determine the correctness of \chatgpt{}'s responses, two authors independently inspected the responses and labeled them. 
For any disagreements, a third author was involved and the questions were discussed so that all three authors came to an agreement.
All three evaluators have either (1) taken a graduate-level software testing course that uses the textbook~\cite{ammann2016introduction} our study questions are sourced from, (2) taught a software testing course that uses the textbook as a teaching assistant, or (3) taught a software testing course that uses the textbook as the instructor.

Responses from \chatgpt{} were broken down into two parts - the \emph{explanation} and \emph{answer}. 
Our labeling considered two perspectives: (1) whether the overall answer was correct\Space{ or not} and (2) whether the explanation given was correct\Space{ or not}.
Both perspectives are categorized as \emph{correct}, \emph{incorrect}, or \emph{partially correct} relative to the solution of the question.

A \chatgpt{} response example to a question is the following:

\begin{boxE}
\textbf{\small Textbook question~\cite[p.114]{ammann2016introduction}}: \small Suppose that coverage criterion C1 subsumes coverage criterion C2.
Further, suppose that test set T1 satisfies C1 on program P, and test set T2 satisfies C2, also on P. Does T1 necessarily satisfy C2? Explain.
\end{boxE}
\begin{boxI}
\textbf{\small \chatgpt{}'s response:} \small T1 may or may not satisfy C2. The fact that T1 satisfies C1, which subsumes C2, means that C1 is a more comprehensive criterion that includes all of the requirements of C2. However, it does not guarantee that T1 will also satisfy C2. T1 may include test cases that are not necessary to satisfy C2, and it may not include test cases that are necessary to satisfy C2. Therefore, additional test cases may be needed to confirm that T1 also satisfies C2.
\end{boxI}

The above example response was labeled as having the answer as \textit{Incorrect}, but the explanation as  \textit{Partially Correct}. 
As \chatgpt{}'s answer is that T1 may \textit{\textbf{not}} satisfy C2 (when the correct answer is that T1 \textit{\textbf{does}} satisfy C2), we label the answer as incorrect. 
The explanation is partially correct, because \chatgpt{} correctly mentioned that C1 includes all of C2's requirements (based on the definition of subsumption in Section~\ref{sec:background:example}), but also incorrectly explains that this fact does not guarantee that T1 satisfies C2.\OurComment{
could not make the logical conclusion that test set T1 should satisfy C2, as T1 satisfies C1 and C1 subsumes C2.}



\section{Results}

\subsection{\RQONE}
\label{sec:eval:rq1}

\subsubsection{\underline{Answer Correctness}} 
\begin{figure}[t]
\vspace{-1em}
\includegraphics{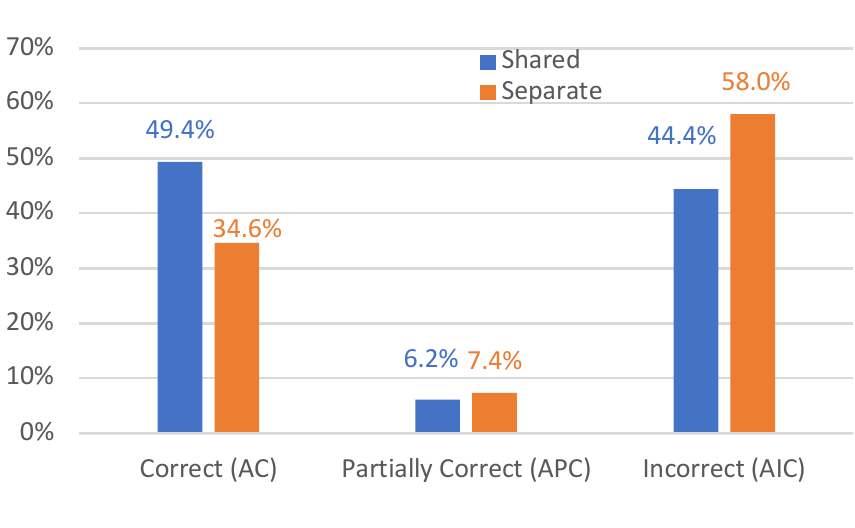}
\prefigcaptionspace
\caption{Correctness of \chatgpt{} answers for shared and separate contexts.}
\label{fig:fig-rq1-ans}
\vspace{-1em}
\end{figure}

We find that in shared contexts, \sharedAnsAcc{}\% of the time the answer is correct, and 6.2\% of the time it is partially correct. 
In contrast, in separate contexts, responses are correct 34.6\% of the time and partially correct 7.4\% of the time. As shown in Figure~\ref{fig:fig-rq1-ans}, a shared context produces fewer incorrect answers than separate contexts, on average. 
One explanation for this behavior is that, in a shared context, \chatgpt{} obtains  contextual information from the prior \subquestion{}s. For example, if \chatgpt{} already identified a fault in a prior \subquestion{}, it is more likely to correctly leverage this fault to answer subsequent questions about errors.

\subsubsection{\underline{Explanation Correctness}} 
\begin{figure}[t]
\includegraphics{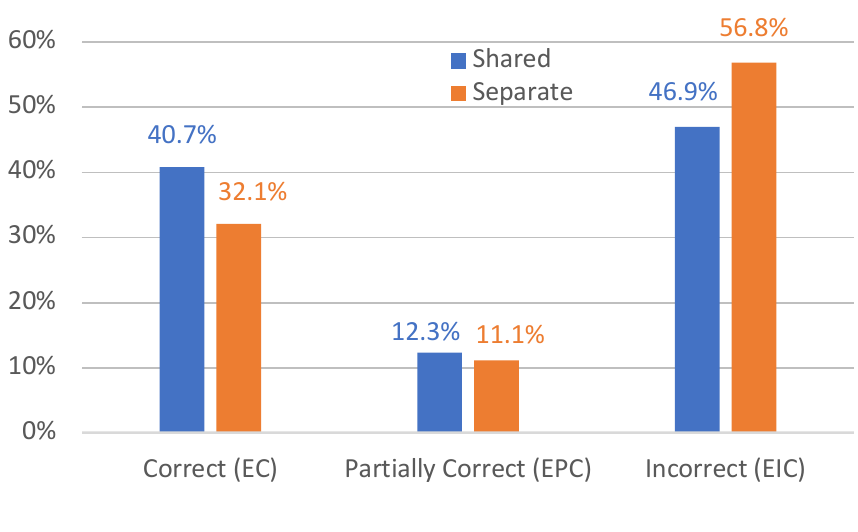}
\prefigcaptionspace
\caption{Correctness of \chatgpt{} explanations for shared and separate contexts.}
\label{fig:fig-rq1-exp}
\vspace{-1em}
\end{figure}

The results for explanation correctness are shown in Figure~\ref{fig:fig-rq1-exp}. 
Here, we obtain similar results to answer correctness. 
Namely, explanation accuracy is higher in the shared context. 
That being said, when we focus on being fully correct in answers and explanations, we find that shared context is better than separate context.
With this finding in mind, our remaining RQs focus solely on our shared context results.
Separate context related data for the remaining RQs is on our website~\cite{studygithub}.

\begin{tcolorbox}
\linespread{0.5} \textbf{\small Shared context is more likely than separate context to be correct. Using \chatgpt{} in a shared context can result in a correct or partially correct answer \answerAccuracyWithPartial{}\% of the time and a correct or partially correct explanation \explanationAccuracyWithPartial{}\% of the time.
}
\end{tcolorbox}

\subsection{\RQTWO}	

\newcommand{\shared}{Sha.} 
\newcommand{\separate}{Sep.} 

Table~\ref{tab:shared_inc} shows results of correctness for answer-explanation pairs across the three iterations (e.g., three responses for each question) for our defined shared query context.  
For example, the Answer Incorrect (AIC) and Explanation Partially Correct (EPC) pair means \chatgpt{} provided an incorrect answer but a partially correct explanation. 
Of the nine possible pairs, there are three pairs where the degree of correctness of the answer and the explanation are the same: (1) AC-EC where both the answer and the explanation are correct, (2) APC-EPC where both are partially correct, or (3) AIC-EIC where both are incorrect.
All other pairs have different degrees of correctness.
Our results suggest that, on average, 
11.8\% of the time, \chatgpt{} is giving an answer that does not properly match its explanation in terms of correctness.

\begin{tcolorbox}
\textbf{\small 11.8\% of the time \chatgpt{} produces responses where the answer-explanation pairs have different degrees of correctness (e.g., the answer is correct, but the explanation is not)\Space{, and the other way around}.\Space{ and it produces identical pairs in the remaining 88.2\% of the time.}}
\end{tcolorbox}	

\begin{table}[t]
\vspace{-1em}
\caption{\label{tab:shared_inc} Answer-Explanation pairs for three shared context iterations. We define three top-level columns for answer correctness: answer correct (AC), answer partially correct (APC), and answer incorrect (AIC), from left to right. For each column under the top level, we show the classification of the explanation for the given top-level answer type:  explanation correct (EC), explanation partially correct (EPC) and explanation incorrect (EIC), from left to right.}
\postcaptionsCodespace
\begin{tabular} { c | r r r | r r r | r r r r}
  & \multicolumn{3}{c|}{\textbf{\underline{AC}}} & \multicolumn{3}{c|}{\textbf{\underline{APC}}} & \multicolumn{3}{c}{\textbf{\underline{AIC}}}\\
 \textbf{Iter.} & \textbf{EC} & \textbf{EPC} & \textbf{EIC} & \textbf{EC} & \textbf{EPC} & \textbf{EIC} & \textbf{EC} & \textbf{EPC} & \textbf{EIC} \\
\hline
 \textbf{1} & 15 &	0 &	2 &	0 &	1 &	0 &	0 &	2 &	11 \\
 \textbf{2} & 15 &	0 &	2 &	0 &	2 &	0 &	0 &	1 &	11 \\
\textbf{3} & 15 &	1 &	2 &	0 &	2 &	0 &	0 &	1 &	10 \\
\hline
\textbf{Sum} & 45 & 1 &	6 & 0 &	5 &	0 &	0 &	4 &	32 \\
\hline
\textbf{\%} & 48.4 & 1.1 &	6.4 & 0.0 &	5.4 &	0.0 &	0.0 &	4.3 &	34.4 \\
\end{tabular}
\vspace{-1em}
\end{table}

\subsection{\RQTHREE}
\Space{
In this RQ\Space{research question}, we analyze how often \chatgpt{} will give responses that differ in correctness for any given question.

For example, if a response has a correct answer, but another response from the next iteration of the same question has an incorrect answer, then the answers for this question are inconsistent.
The same goes for explanations.
We find that, of the \numAllQuestions{} questions we study, \chatgpt{} gives inconsistent answers for 9.7\% of the questions and inconsistent explanations for 6.5\% of the questions. In this RQ\Space{research question}, we analyze how often \chatgpt{} will give responses that differ in correctness for any given question.
For example, if a response has a correct answer, but another response from the next iteration of the same question has an incorrect answer, then the answers for this question are inconsistent.
The same goes for explanations.
We find that, of the \numAllQuestions{} questions we study, \chatgpt{} gives inconsistent answers for 9.7\% of the questions and inconsistent explanations for 6.5\% of the questions.
}

When asked the same question multiple times, \chatgpt{} may give a different response each time due to the stochastic nature of its sampling process from a learned probability distribution. 
We examine how often these differing responses given by \chatgpt{} will differ in correctness.
For example, a question may have a correct answer in one run but an incorrect answer in another run. 
We find that for 9.7\% of questions, the answer's correctness is affected by non-determinism and for 6.5\% of questions, the explanation's correctness is affected.

\begin{tcolorbox}
\textbf{\small The correctness of \chatgpt{}'s answers vary between correct to incorrect for 9.7\% of questions, and the correctness of explanations varies for 6.5\% of questions.}
\end{tcolorbox}

\subsection{\RQFOUR}

\begin{figure}[t]
\vspace{-1em}
\includegraphics{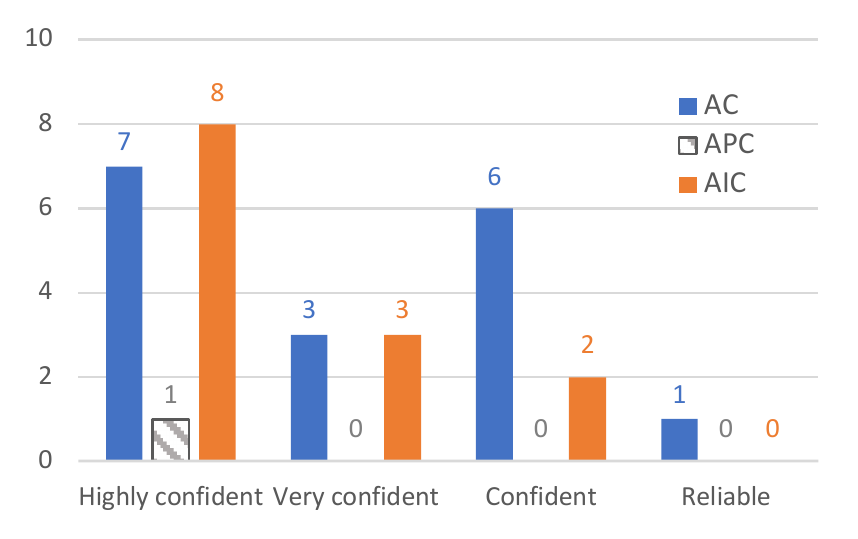}
\precaptionspace
\vspace{-1.5ex}
\caption{ChatGPT's reported confidence for correct, partially correct, and incorrect answers.}
\label{fig-rq4-ans}
\end{figure}

\begin{figure}[t]
\vspace{-1em}
\includegraphics{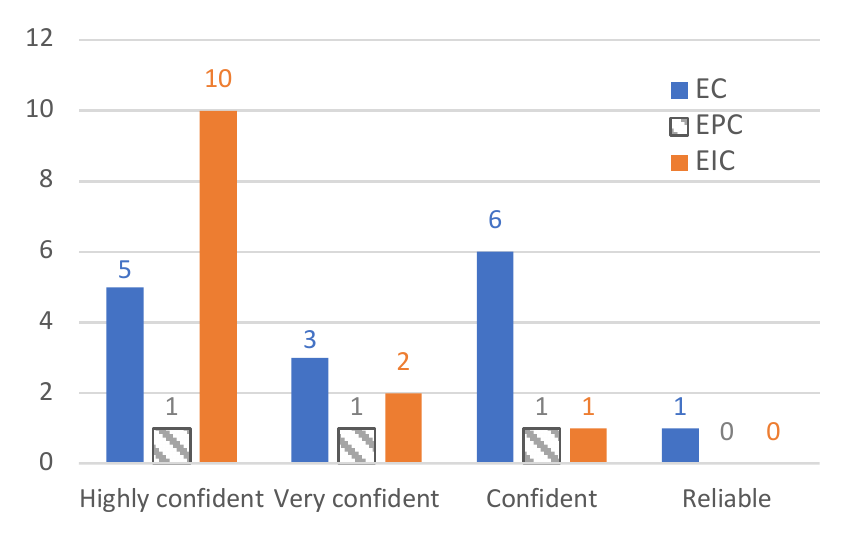}
\precaptionspace
\vspace{-1.5ex}
\caption{ChatGPT's reported confidence for correct, partially correct, and incorrect explanations.}
\label{fig-rq4-exp}
\vspace{-1em}
\end{figure}

To determine how confident \chatgpt{} is in its answers, we asked it to report its confidence. 
Asking about confidence is typically referred to as ``calibration'', and a well-calibrated model will be confident when correct, and less confident when incorrect. In our experiment, \chatgpt{} responded with four different keywords. 
Figures~\ref{fig-rq4-ans} and~\ref{fig-rq4-exp} display the\Space{ corresponding} data for each keyword for answers and explanations, respectively.

\chatgpt{} expressed varying levels of confidence in its responses. 
When \chatgpt{} is ``Highly confident'' in its response, we find that its answers are incorrect about half the time and explanations are incorrect twice as often as correct. On the other hand, when \chatgpt{} is ``Confident'' in its response, we find that its answers or explanations are at least three times as likely to be correct than incorrect. 
For the other two categories, we find that \chatgpt{} provides mixed responses. 

\begin{tcolorbox}
\textbf{\small \chatgpt{}s self-reported confidence does not appear to be particularly useful, as it has little bearing on question correctness. This finding seems to indicate, that, for software testing questions, \chatgpt{} is not well calibrated.}
\end{tcolorbox}

\begin{figure}[t!]
\begin{center}  
\includegraphics[scale=0.30]{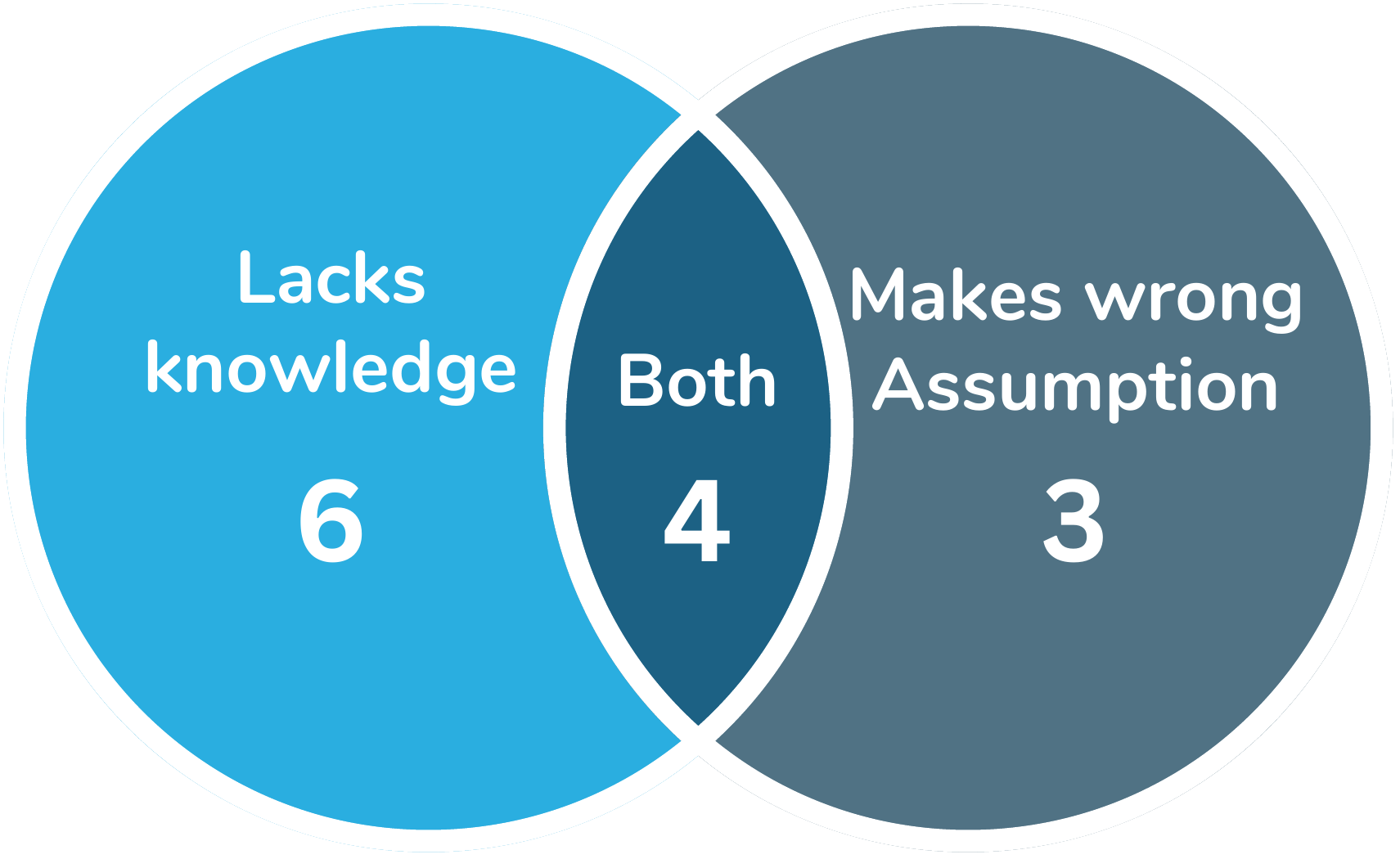}
\vspace{-1em}
\caption{Venn diagram of characteristics of 13 incorrect answers from \chatgpt{}.}
\label{fig:fig-venn}
\end{center}
\vspace{-2ex}
\end{figure}

\section{Case Study}
\label{sec:discussion:case}


In this section, we present a case study examining (1)~the characteristics of incorrect answers, (2)~how \chatgpt{}'s responses may change when we provide it with more information, and (3)~an example of an inconsistent answer-explanation pair. 
The goal of our case study is to gather insight into why \chatgpt{} is wrong and to examine how certain prompting strategies may lead to a higher likelihood of correct answers.

\subsection{Characteristics of Incorrect Answers from \chatgpt{}}
\label{sec:study:characteristics}
To better understand why \chatgpt{} is wrong, we categorized 13 incorrect answers (out of 31) from shared context's first iteration, and identified three main reasons for \chatgpt{}'s incorrect responses.
Figure~\ref{fig:fig-venn} summarizes our findings.

\subsubsection{\chatgpt{} lacks knowledge} The first category is when \chatgpt{} may lack the knowledge needed to solve the problems given to it. For questions from chapters 1, 2, and 3, \chatgpt{} seemed to lack definitions for fault, failure, and error, resulting in incorrectly treating errors as failures or crashes.
For Chapter 5, \chatgpt{} seemed to lack understanding of how to conclude whether a test set satisfies a coverage criterion. 
An example of \chatgpt{} getting the answer incorrect due to lack of knowledge is presented in Section~\ref{sec:lackinfo:example}.

\subsubsection{\chatgpt{} makes the wrong assumption} \chatgpt{} can also focus on an irrelevant part of the question and give an incorrect answer because it makes an incorrect assumption about what is important. See Section~\ref{sec:discussion:case:moreinfo} for an example.

\OurComment{
These include- 
\begin{itemize}
    \item Focuses on NullPointerException (NPE), when the fault was a logical mistake.
    \item Asks to change Java TreeSet implementation whereas the actual fault was in the \CodeIn{equals()} of HashSet.
    \item Provides an incorrect solution using inheritance, while the actual solution should use composition.
\end{itemize}
}

\subsubsection{Both} In four cases, \chatgpt{} seems to both lack knowledge and make wrong assumptions.
For example, for one of the four cases, \chatgpt{} makes a wrong assumption about the fault in a program and lacks the knowledge for what is an error. 
Therefore, \chatgpt{}'s response about the error in this program was incorrect due to both characteristics.\Space{
For \subquestion{}s in a shared context, a wrong assumption or lack of knowledge for one question can affect all subsequent questions.}

\OurComment{
Some of the mistakes made by \chatgpt{} because of lack of knowledge are- 

\begin{itemize}
\item Treats error state as a failure
\item Treats error state as exception/crash
\item Treats crash/exceptions as failure
\item Knows what subsumption is but cannot make a logical decision
\end{itemize}
}

\subsection{Effect of Additional Information on \chatgpt{}'s responses}
\label{sec:discussion:case:moreinfo}
To understand the effect of additional information on \chatgpt{}, we randomly select one of the 13 incorrect answers mentioned in Section~\ref{sec:study:characteristics} and manipulate the prompt\Space{ in an effort}\Space{ to get \chatgpt{} to produce the correct response}.

\begin{boxE}
\textbf{\small Textbook question~\cite[p.41]{ammann2016introduction}:}
\label{code_oddpos}
\begin{lstlisting}[language=Java,basicstyle=\ttfamily\scriptsize,escapeinside={(*@}{@*)},columns=fixed,xleftmargin=3.5ex]
public static int oddOrPos (int[] x) {
  int count = 0;
  for (int i = 0; i < x.length; i++) {
    if (x[i]%2 == 1 || x[i] > 0) count++;
  }
  return count;
} 
// test: x = [-3,-2,0,1,4]; Expected = 3
\end{lstlisting}
\small \textbf{(a)} Explain what is wrong with the given code. Describe the fault precisely by proposing a modification to the code. 
\end{boxE}

The function \oddOrPos{} from the textbook (shown above) is about finding odd or positive numbers. But the given code snippet fails to count negative odd values because a negative odd number produces \CodeIn{-1} when it is \CodeIn{mod} with \CodeIn{2}. To answer this question, \chatgpt{} should find the fault and show a possible modification to the program.  
\chatgpt{} originally thought the fault is related to not having a null check at the beginning for the input of a null array. But the actual fault is that it does not take into account negative odd values.
In its response, \chatgpt{} advises adding a null check before the \CodeIn{for} loop, which is an incorrect fix.

\begin{boxE}
\textbf{\small Modified prompt:} 
\begin{lstlisting}[language=Java,basicstyle=\ttfamily\scriptsize,escapeinside={(*@}{@*)},columns=fixed,xleftmargin=3.5ex]
public static int oddOrPos (int[] x) { ... }

\end{lstlisting}

\small \textbf{(a)} ... The fault and the modification do not involve having a null check.

\end{boxE}
We then changed the \subquestion{} (a) to provide more information (see above). The modified prompt mentions that the fault and the modification are not related to a null check. 
\OurComment{
\Fix{Wing: We must have also done this for the original prompt then right? More importantly, we may just want to pick a simpler example that isn't the fifth/sixth subquestion in a shared context.}
According to our findings from RQ1, \chatgpt{} performs better in a shared context, therefore, we provide \chatgpt{} with the faulty code snippet and all of the \subquestion{}s.
}
The textbook solution refers to changing the \CodeIn{if} condition to \CodeIn{x[i]\%2 == -1 || x[i] > 0}, where the negative odd numbers are processed by the first part of the expression.
With these modifications to the prompt, ChatGPT’s response is
correct (see below): it checks for both positive and negative odd numbers
with \CodeIn{x[i]\%2 != 0}.

\begin{boxI}
\textbf{\small \chatgpt{}'s response to modified prompt:} 
\small The given code has a logical error. The condition x[i]\%2 == 1 checks for odd elements but does not include negative odd numbers. 
To repair the code, we can change the if statement to check if for odd 
\CodeIn{(x[i] \% 2 != 0)} or positive \CodeIn{(x[i] > 0)}.
\begin{lstlisting}[language=Java,basicstyle=\ttfamily\scriptsize,escapeinside={(*@}{@*)},columns=fixed,xleftmargin=3.5ex]
public static int oddOrPos (int[] x) {
  ...
  if (x[i]%2 != 0 || x[i] > 0)
  ...
}
\end{lstlisting}
\vspace{-2ex}
\end{boxI}

\subsection{Answer Correct, Explanation Incorrect (AC-EIC) Example}
In Section~\ref{sec:lackinfo:example}, we showed an example of AIC-EPC. 
In this section, we present an opposite case -- an example where the answer is correct, but the explanation is incorrect.
This example is another \subquestion{} of the \CodeIn{oddOrPos} code presented in Section~\ref{sec:discussion:case:moreinfo}.
\begin{boxE}
\small \textbf{(d)} If possible, give a test case that results in an error, but not a failure. If not, briefly explain why not. Hint: Don’t forget about the program counter.
\end{boxE}

\begin{boxI}
\textbf{\small \chatgpt{}'s response:}
\small It is not possible to give a test case that results in an error, but not a failure because when the code throws a \CodeIn{NullPointerException}, it results in an error and the program will stop execution. An error state refers to the unexpected behavior of the program. The program counter will not continue execution past the point where the exception is thrown.
\end{boxI}

According to the textbook, every input that results in an error also results in a failure, because the error states can not be repaired\Space{ by subsequent processing if there is a negative value in \CodeIn{x}}. 
The answer from \chatgpt{} is correct because it is impossible to give a test case that results in an error but not a failure. 
However, the explanation is incorrect as \chatgpt{} treats the \CodeIn{NullPointerException} as the reason.

\section{Threats to Validity}


One threat to the validity of our study is the dataset we used. 
The exercises may not address all the domains required for a software testing class. 
To address this limitation, we selected a book~\cite{ammann2016introduction} that is widely used and included all questions with student solutions~\cite{booksolution}\Space{ contains key concepts found in most software testing books} from the first five chapters.

As \chatgpt{} was in research preview at the time of this study and is continuously updated, its behavior may differ in future iterations. We identified a few improvements to its performance in the course of our study. In earlier releases, \chatgpt{} was unable to provide a numeric confidence level, while it is now able to specify a level between 0.0 to 1.0.
Responses from \chatgpt{} are also inconsistent, where repeated invocations of \chatgpt{} with the same question yield different responses. 
To reduce the effect of inconsistency on our study, we ran each question three times for generalizability.

Finally, our main results made limited use of prompt engineering (i.e., only varying question context), where differently designed prompts might yield more correct answers. Except for our case study (Section~\ref{sec:discussion:case}), our results are based on responses obtained by asking the questions directly as they appear in the book.
Book practice questions are designed to focus on human readers and are usually based on the contents of a corresponding chapter. 
As we have not provided \chatgpt{} with the actual contents of the chapters, it is possible that \chatgpt{} might be correct more often if additional context is provided with the questions.

\section{Related Work}

Several systems have been proposed to apply large language models (LLMs) to the problem of generating code snippets from natural language requests from developers~\cite{Li2022AlphaCode}. Most prominently, GitHub CoPilot, based on Codex, popularized the use of LLMs for real-world programming tasks~\cite{Codex2021}. 

Studies have begun to examine how effectively these systems may be used for code generation tasks. AlphaCode was found to generate code that was often similar to human-generated code~\cite{AlphaCodeEval2022} and achieved a simulated average ranking in the top 54\% on Codeforces~\cite{Li2022AlphaCode}, a programming competition platform.
One study found that, on tasks to fix security defects after the defect has already been localized and where additional information is provided through the prompt, LLMs can successfully generate fixes~\cite{LLMsecurityfixes2023}.   

Some work has specifically examined the potential use of LLM code generation by students in computer science courses. One study found that Codex already performs better than most students on the code writing questions found in typical\Space{ first-year} introductory programming exams~\cite{Finnie22-CodexIntroProg} as well as more advanced exams on data structures and algorithms~\cite{Finnie23-CodexDatastructures}. Identical prompts frequently lead to widely varying algorithm choices and code size~\cite{Finnie22-CodexIntroProg}. A study examining CoPilot's performance on programming assignments from introductory courses found that it achieved scores of\Space{ between} 68\% to 95\%~\cite{Puryear22-CoPilotClassroom}.

However, there are substantial challenges with the usability of these systems that may limit their effectiveness for real-world programming tasks. One study found that, despite participants themselves enjoying interacting with the code generation system, there was no measurable productivity benefit in either speed or correctness of programming tasks~\cite{Xu2022-CodeGenStudy}. Similarly, a second user study found that the productivity benefits of CoPilot were mixed: while it sometimes could make developers faster, it could also lead developers down time-consuming rabbit holes debugging incorrect code~\cite{VaithilingamCHI2022-CoPilotStudy}. 
As a result,\Space{ overall,} it had no significant impact on the correctness or task time. 
\Space{Examining the security implications of using these systems, one study found that developers using an AI code assistant were more likely to believe that they wrote secure code but had in fact written less secure code\Space{ that was significantly less secure} than if they had not used the AI code assistant~\cite{StanfordInsecureCodeStudy}.}
Including explanations may help reduce overreliance on potentially incorrect answers, but only in situations when the benefits of engaging with explanations outweigh the costs~\cite{Vasconcelos2023-ReduceOverreliance}.

As \chatgpt{} was only recently released, there are only a few studies that have specifically examined the effectiveness of \chatgpt{} on various tasks. Investigations have found that \chatgpt{} produces responses that are at or near the passing threshold for all three parts of the US Medical Licensing Exam without any additional information or prompt engineering~\cite{Kung2022-USMLEEval}. 
ChatGPT has also been able to achieve a low but passing average grade of C+ in four law school classes~\cite{Choi2021-ChatGPTLawSchool}.
\Space{\chatgpt{} also may lower the barriers for threat actors in the security domain, helping threat actors with little experience to create malware, social engineering attacks, disinformation, and phishing~\cite{IChatBot}.  These new capabilities have already been used by threat actors to share buggy but functional new attacks on dark web forums.}

\section{Discussion: Promises \& Perils}

In this paper, we examined the potential applicability of \chatgpt{} to a popular software testing curriculum. We found that 
\chatgpt{} is able to provide correct or partially correct answers to \answerAccuracyWithPartial{}\% of questions. 
Moreover, \chatgpt{} is a poor judge of its own correctness: its confidence has little bearing on the correctness of its response\OurComment{ more likely to have a correct answer when it is only confident rather than highly confident in its answer}. 
Said differently, at least when this study was conducted, \chatgpt{}'s answers will, more likely than not, be incorrect for questions related to software testing courses. 
That being said, our findings still raise immediate concerns\Space{, as well as\Space{ a number of} questions} on how the use of \chatgpt{} might be detected to ensure that questions are meaningfully assessing students' understanding of course materials.

\OurComment{
However, we also found that \Space{ overall,} \chatgpt{} is not able, by itself, to pass a software testing course. \Fix{Don't talk about passing or not. There is no formal definition of what percentage correct is needed to ``pass''. The only thing our results show is that \chatgpt{} is unable to respond with the correct answer for the majority of the questions in our study.} 
}

While concerns over student's use of \chatgpt{} to circumvent assessments represent one potential \textit{peril}\Space{ related to the use of \chatgpt{}}, there are also several \textit{promising} directions for integrating \chatgpt{} into the classroom. 
We found that \chatgpt{} is able to provide correct or partially correct explanations to \explanationAccuracyWithPartial{}\% of questions. 
Furthermore, we found that using certain prompting strategies, which provide additional question context, can improve the chances of correct answers and explanations. 
This finding suggests that for carefully designed in-class activities or labs, \chatgpt{}, rather than an instructor or TA, can be used to guide students through a set of exercises to improve students' understanding of the material\Space{, if the activity is carefully crafted}. 
 
Furthermore, we found that certain contexts make it difficult for \chatgpt{} to answer correctly, and such settings could be used to prevent cheating, especially when access to the internet is necessary.
Our dataset contains coding and conceptual questions, and some questions that are both.
In our experiments, \chatgpt{} performed worst with questions involving both code and concepts. 
It outputs correct answers and explanations most often with coding questions (83.3\%), then with conceptual questions (55.6\%), and finally with combined questions~(31.3\%).

Our results are in contrast to the results of applying \chatgpt{} in other domains, such as in medicine~\cite{Kung2022-USMLEEval} or law~\cite{Choi2021-ChatGPTLawSchool} where \chatgpt{} is shown to pass certain parts of their exams. 
This difference may be due to the fact that there may be far more content available with which \chatgpt{} may be trained for these exams, or perhaps due to differences in the nature of the questions themselves. 
As \chatgpt{} has only recently been released, a full picture of its capabilities and the impact of such tools is still yet to be determined.

\section*{Acknowledgement}
\noindent We thank Abdulrahman Alshammari, Paul Ammann, Talank Baral, Atish Dipongkor, Safwat Ali Khan, Ajay Krishnavajjala, Mikael Lindvall, Jeff Offutt, and Adam Porter for their feedback on this work.

\bibliographystyle{ieeetr}
\bibliography{main}

\end{document}

%% file: macros.tex
\newcommand{\Fix}[1]{{\color{red}\bfseries[[#1]]}}

\newcommand{\OurComment}[1]{}
\newcommand{\Space}[1]{}
\newcommand{\CodeIn}[1]{{\small\texttt{#1}}}

\newcommand{\Def}[2]{\expandafter\newcommand\csname rmk-#1\endcsname{#2}}
\newcommand{\Use}[1]{\csname rmk-#1\endcsname}

\Def{sleepyline.cause}{Sleep Reasons}

\Def{config.Count}{Configs. (\%)}
\Def{configs.proportion}{configs.proportion}

\Def{failureID.Uniq.Count}{Failures (\%)}
\Def{failureID.Uniq.Proportion}{failures.proportion}

\Def{test.count}{Tests (\%)}
\Def{test.proportion}{tests.proportion}

\Def{FLAGGED_API}{API}
\Def{RUNNABLE_CALLABLE_START}{ThreadBegin}
\Def{MONITORENTER}{EnterSyncBlock}
\Def{MONITOREXIT}{ExitSyncBlock}
\Def{SYNCHRONIZED_METHOD_ENTER}{EnterSyncMethod}
\Def{SYNCHRONIZED_METHOD_EXIT}{ExitSyncMethod}
\Def{java/lang/System.currentTimeMillis()J}{System.currentTimeMillis}
\Def{java/lang/Thread.currentThread()Ljava/lang/Thread;}{Thread.currentThread}
\Def{java/lang/Object.notifyAll()V}{Object.notifyAll}
\Def{java/io/OutputStream.flush()V}{OutputStream.flush}
\Def{java/nio/ByteBuffer.allocate(I)Ljava/nio/ByteBuffer;}{ByteBuffer.allocate}
\Def{java/nio/ByteBuffer.array()[B}{ByteBuffer.array}
\Def{java/nio/channels/Selector.wakeup()Ljava/nio/channels/Selector;}{Selector.wakeup}

\newcommand{\prefigcaptionspace}{\vspace{-5ex}} 
\newcommand{\precaptionspace}{\vspace{-2ex}}

\newcommand{\postcaptionsCodespace}{\vspace{-2ex}} 

\newboolean{showcomments}

\setboolean{showcomments}{true}

\ifthenelse{\boolean{showcomments}}
  {\newcommand{\nb}[2]{
    \fbox{\bfseries\sffamily\scriptsize#1}
    {\sf\small$\blacktriangleright$\textit{#2}$\blacktriangleleft$}
   }
   
  }
  {\newcommand{\nb}[2]{}
   
  }

\definecolor{Green}{rgb}{0.6,1,0.8}
\definecolor{Red}{rgb}{0.99, 0.76, 0.8}

\definecolor{dkgreen}{rgb}{0,0.6,0}
\definecolor{gray}{rgb}{0.5,0.5,0.5}
\definecolor{mauve}{rgb}{0.58,0,0.82}

\newcommand\KEVIN[1]{{\color{blue} \nb{KEVIN}{#1}}}

\newcommand\RQONE{\textbf{RQ}$_1$: How often is \chatgpt{} able to provide correct answers and explanations for different prompting strategies?}
\newcommand\RQTWO{\textbf{RQ}$_2$: How often does \chatgpt{} give answer-explanation pairs with different degrees of correctness?}
\newcommand\RQTHREE{\textbf{RQ}$_3$: How does \chatgpt{}'s non-determinism affect its ability to provide correct answers and explanations?}
\newcommand\RQFOUR{\textbf{RQ}$_4$: How does \chatgpt{}'s confidence in its response correlate to the correctness of the response?}

\newcommand\AO{Ammann and Offutt}
\newcommand\chatgpt{ChatGPT}

\newcommand\numSubQuestions{27}
\newcommand\numAllQuestions{31}
\newcommand\numAllQuestionsThreeRuns{93}
\newcommand\numNonSubQuestions{four}

\newcommand\numQuestionsWSub{six}

\newcommand\tenSeconds{10}

\newcommand\subquestion{sub-question}
\newcommand\lastZero{\CodeIn{lastZero()}}
\newcommand\oddOrPos{\CodeIn{oddOrPos()}}
\newcommand\answerAccuracyWithPartial{55.6}
\newcommand\explanationAccuracyWithPartial{53.0}
\newcommand\sharedAnsAcc{49.4}

\newcommand\confidenceQuestion{How confident are you that your previous response is correct?}
\newcommand\codeQuestion{6}
\newcommand\conceptualQuestion{9}
\newcommand\combinedQuestion{16}

\definecolor{main}{HTML}{5989cf}    
\definecolor{sub}{HTML}{cde4ff}     
\newtcolorbox{boxI}{
    colback = sub, 
    colframe = main, 
    boxrule = 0pt, 
    toprule = 4pt 
}

\newtcolorbox{boxE}{
    colframe = main, 
    boxrule = 0pt, 
    toprule = 4pt 
}

\tcbset{tab2/.style={enhanced,fonttitle=\bfseries,fontupper=\normalsize\sffamily,
colback=yellow!10!white,colframe=red!50!black,colbacktitle=Salmon!40!white,
coltitle=black,center title}}

\def\BibTeX{{\rm B\kern-.05em{\sc i\kern-.025em b}\kern-.08em
    T\kern-.1667em\lower.7ex\hbox{E}\kern-.125emX}}

\makeatletter
\newcommand{\linebreakand}{%
  \end{@IEEEauthorhalign}
  \hfill\mbox{}\par
  \mbox{}\hfill\begin{@IEEEauthorhalign}
}